# Grid Computing in the Collider Detector at Fermilab (CDF) scientific experiment

D. Benjamin (for the CDF Collaboration)

*Duke University, Durham, NC 27708, USA*

The computing model for the Collider Detector at Fermilab (CDF) scientific experiment has evolved since the beginning of the experiment. Initially CDF computing was comprised of dedicated resources located in computer farms around the world. With the wide spread acceptance of grid computing in High Energy Physics, CDF computing has migrated to using grid computing extensively. CDF uses computing grids around the world. Each computing grid has required different solutions. The use of portals as interfaces to the collaboration computing resources has proven to be an extremely useful technique allowing the CDF physicists transparently migrate from using dedicated computer farm to using computing located in grid farms often away from Fermilab. Grid computing at CDF continues to evolve as the grid standards and practices change.

## 1. INTRODUCTION

The Collider Detector at Fermilab (CDF) scientific experiment is a large multipurpose particle physics experiment located at the Fermi National Accelerator Laboratory (Fermilab), west of Chicago Illinois in the United States. The CDF detector measures the particles produced in proton – anti proton collisions at the Tevatron operating at $\sqrt{s}$ = 1.96 TeV The are two experiments at the Tevatron and this paper will describe the grid computing for CDF.

The CDF detector is comprised of a tracking system located closest to the collision point inside of a solenoid magnet. Outside of the solenoid magnet, calorimeter detectors used to measure the energy of the particles resulting from the collisions. Finally on the outside are muon detectors. The CDF experiment started collecting data in 1988 and will continue to take data until at least October 2009. There are ¾ million electronic channels readout and there are collisions every 396 ns.

The CDF computing model has several components. The events passing the trigger are readout and written to robotic tape storage for further calibration, reconstruction and analysis. The reconstruction program reads the data from tape using dCache system [1] to stage the data to disk prior to copying the data files to the worker nodes located in an onsite computer farm and writing the results to tape. CDF users typically analyze the data at FNAL and use remote computer farms for Monte Carlo production with the results send back to FNAL for archival storage.

The CDF analysis farm (CAF) [2] is a software infrastructure developed to allow CDF to use computer farms of commodity hardware for data reconstruction, data analysis, simulated data production. Users develop software, debug programs and submit jobs from their own desktops. User authentication is based on Kerberos v5. Monitoring of system performance and user jobs can be done either through a web interface or command line tools. The interactive monitoring tools allow the users to check the progress of their running jobs by viewing the log files on the worker nodes and obtain other information from the worker nodes (like directory of user job's remote working area). Users can also stop all of part of their jobs if needed. Most of the monitoring functionality is available to the CDF users running on the grid. The users are notified of the completion of their job via job summary sent via e-mail. The user can also specify the location to which output compressed tar ball containing the job logs and whatever files existed at the job completion in the users work area. Kerberized rcp is used to transfer the files at present.

The current version of the CDF CAF used for processing the data with the production reconstruction executable at FNAL is based on a dedicated Condor pool [3]. Condor daemons (collector, negotiator, multiple schedd) run on the head node (or portal machine). The condor starter and startd daemons run on the worker nodes. CAF code includes the job submission client run on the user's desktop. CAF code used on the portal machine includes job server daemon, monitoring daemons (web and command line), mailer daemons. On the worker nodes a CDF specific job wrapper containing the user job executables is used. The initial implementation was not grid complaint and used remote dedicated computer farms. It has since been adapted to use computer farms located at grid sites. Since the users interact to the CDF computing through portals, the users are removed from the underlying computing farm architecture.

## 2. TRANSITION TO GRID COMPUTING

Use of the grid, allowed CDF to reduce the need for dedicated computer farms. As CDF transitioned from dedicated computing farms to using OGS and EGEE grid resources, the number of portals and dedicated computing farms was reduced from 12 dedicated computer farms in 2001-2003 to five portals accessing grid resources in 2006. CDF uses different portals to access computing resources in three different regions of the world (Europe, Asia and North America).

There are numerous requirements for the grid transition of CDF computing. Most importantly, the impact on the CDF users must be minimized. This allows the users to focus on the science and not the computing system. The changes to the CDF middleware must be small. Users are still required to use Kerberos V5 authentication. X509 authentication is used for the grid components. The users must be able to access all of the experiment data (collision and simulated) located at FNAL and large fraction of the data at the CNAF farm in Italy. User job submission and job access must be centralized via the various computing portals. Monte Carlo data production should occur on the grids available to CDF (Open Science Grid (OGS) [4] and Enabling Grids for E-sciencE (EGEE) [5]).

In order to adapt the CDF CAF model to the grid several items needed to be solved. The job submission mechanism needed to be changed from using dedicated computer farms to using computers in grid pools. The job execution environment had to be changed from a CDF specific environment (namely how worker nodes were configured) to a more general grid configuration. OSG and EGEE have similar but different job execution environments. Code distribution and remote database access is different in a grid-computing environment. Each grid (OSG and EGEE) has different mechanisms for monitoring user job status. The user job output retrieval is different in a grid environment than when using a dedicated computer farm. The Fermigrid portal is used to access the CDF computing resources on the FNAL campus grid. NamCAF is used the access the North American OSG grid resources. PacCAF is used to access Asian resources (both EGEE and OSG). LcgCAF is used to access European resources with the EGEE/gLite middleware [6].

## 3. CDF GRID SOLUTIONS

CDF has chosen to use different solutions on each of the grids (OSG and EGEE) that it uses. On the Open Science Grid, Condor glidein's are used to solve the job submission problem. Additional daemons (glidekeeper and glidein schedd) are added to the portal head nodes to manage the glidein's. These glidein's are pilot jobs that contain a Condor startd daemon, which are submitted to the various grid pools. Once Condor startd daemon starts on the worker node, the batch system on the CDF portal sees the worker node as an available batch slot. After the slot is matched, a user job section is transferred to the worker node for execution and the results are sent back the portal head node when the job section is complete. Wherever possible, SQUID caches [7] are used at the grid sites to reduce the load on the CDF portal machine. Job and user tar balls are stored in the Squid caches located near the worker nodes to efficiently deliver the files to the worker nodes. Figure 1 shows how the CDF CAF is used on the OSG including the FNAL campus grid. When sending jobs to the OSG, CDF solves the code distribution issues by sending the software in self contained tar balls. Figure 2 shows how the LcgCAF portal used to send jobs across the EGEE grid using the daemon processes written specifically for this system. This system uses the EGEE workload management system (WMS) for job scheduling. Job monitoring is done through a process that is started along with the user job on the worker node. The monitoring processes talk with the worker node collector daemon running on the portal head node to collect the monitoring information and make it available to user. The CDF software environment is made available on the worker node through the Parrot virtual file system [7]. As with the computing on the OSG, Squid proxy caches are used whenever possible. Figure 2 shows the EGEE portal configuration.

The program Parrot is used to setup a virtual file-system and access CDF software on the worker nodes. Parrot traps the program systems calls and retrieve the needed files. Parrot uses the http protocol for easy caching near the bigger sites by using Squid proxy caches. The CDF software is exported via an Apache web server at CNAF. The programs author – Douglas Thain (University of Notre Dame) provides CDF with excellent support. The use of Parrot on the EGEE grid means that there are no CDF specific software requirements on the worker nodes and only the software needed by the user job is delivered to worker node. The easy caching with the Squid proxy caches improves performance. CDF has plans to distribute CDF code on the OSG with Parrot also. Figure 3 shows how CDF software is distributed to the worker nodes at the various sites in the EGEE using Parrot and Squid Caches.

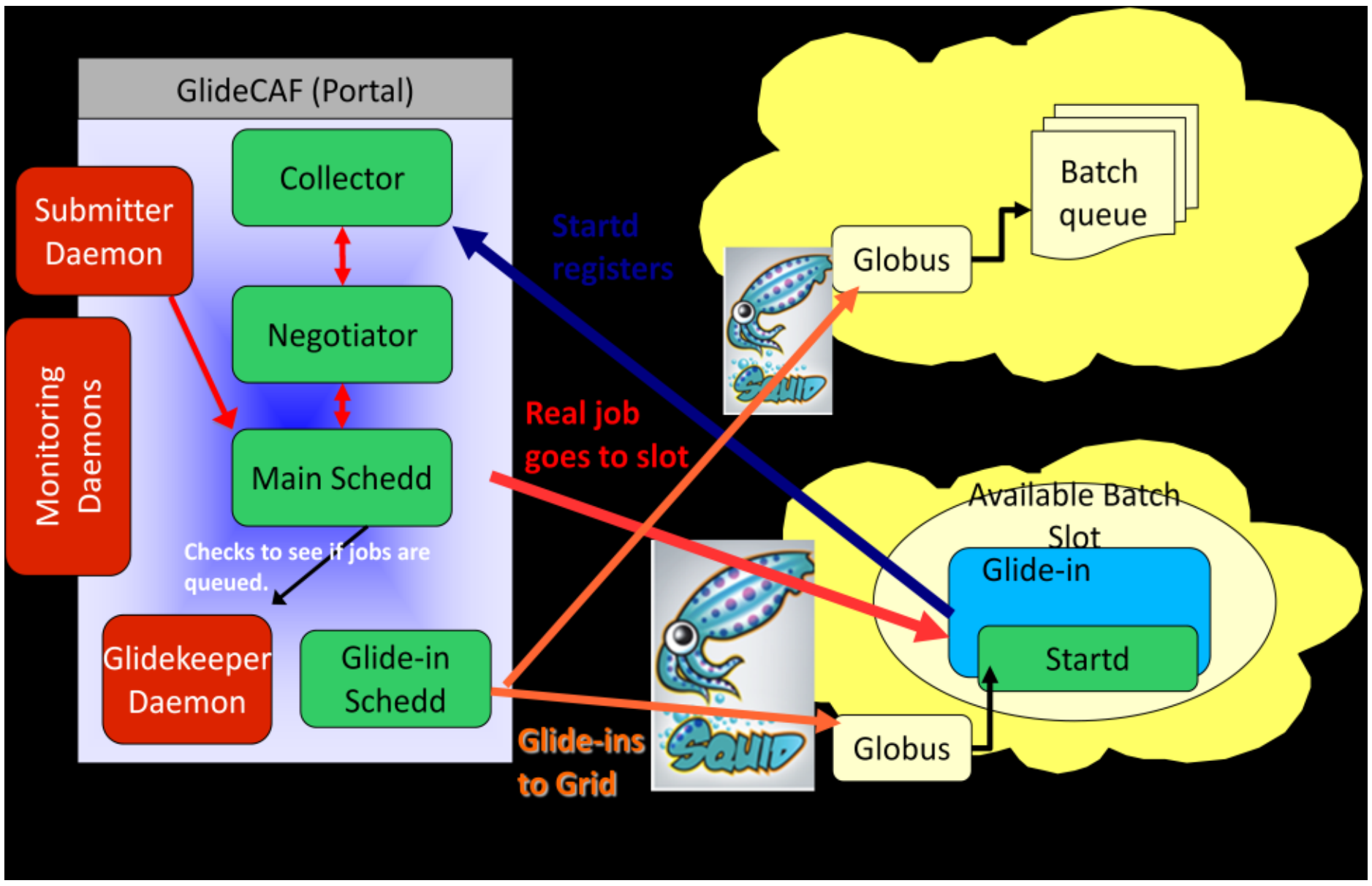

Figure 1 - OSG (including FNAL Campus grid) portal including processes used to send CDF sends jobs to the respective grids. Further details are included in the text.

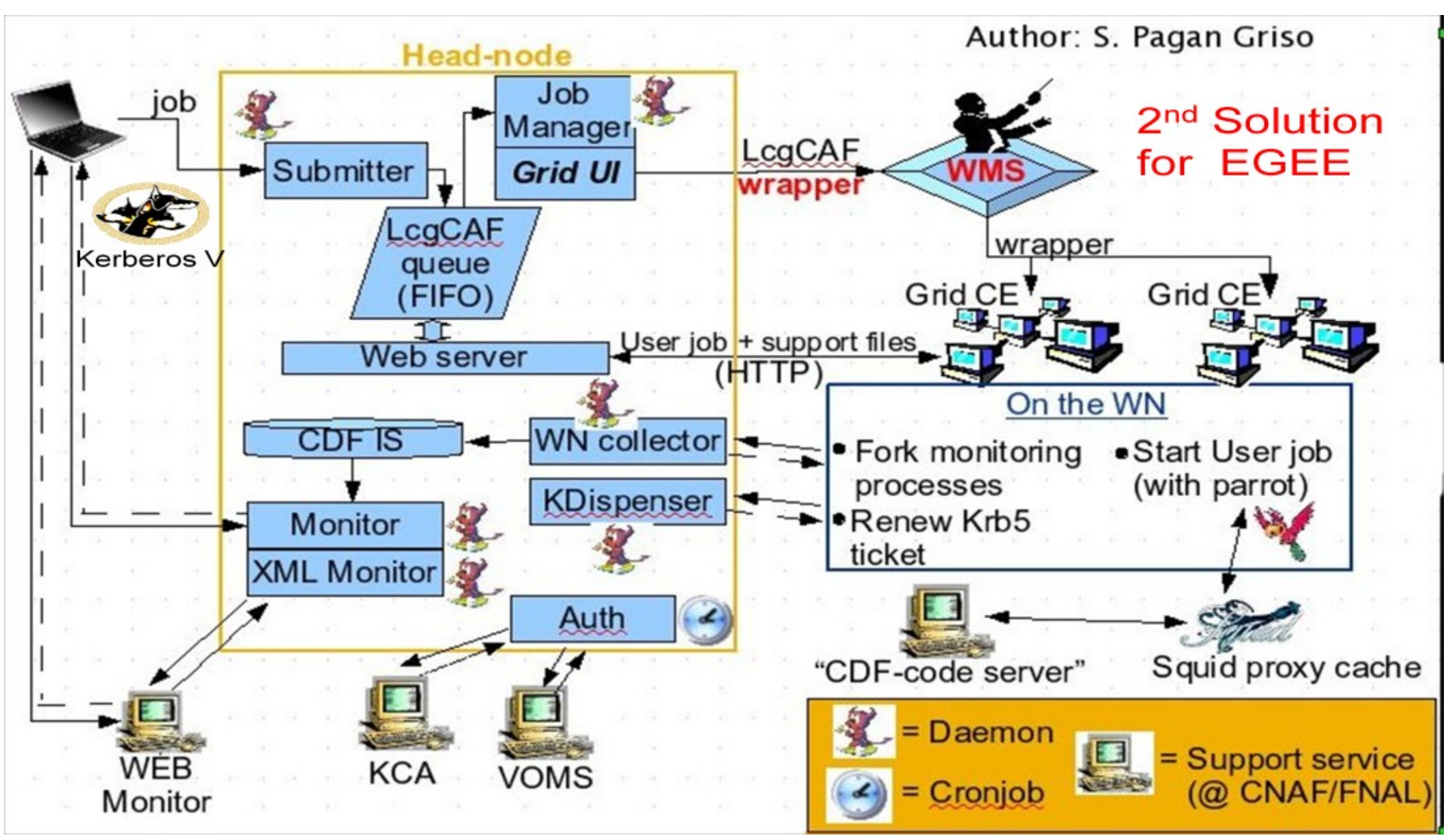

Figure 2 - EGEE portal including processes used in sending CDF jobs to the EGEE. Further details are included in the text.

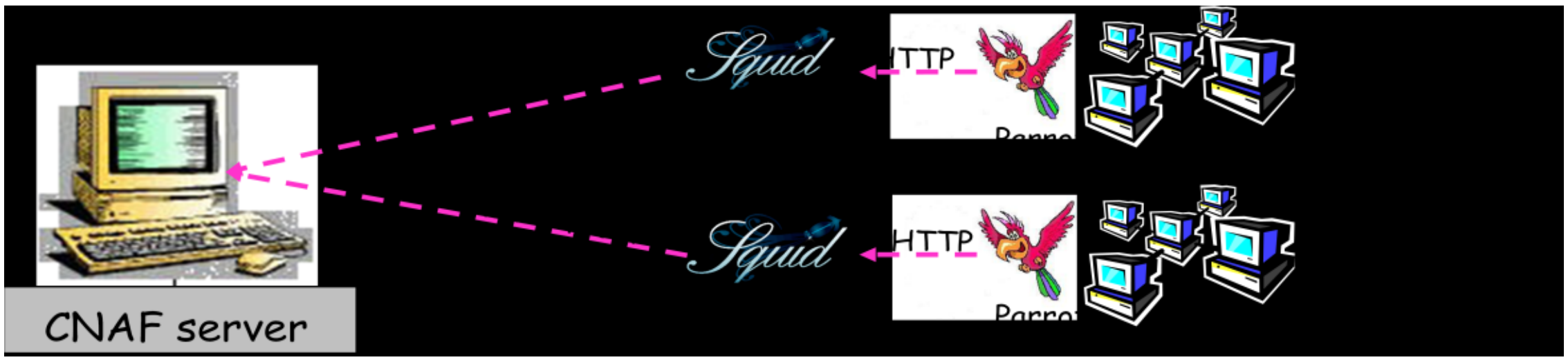

Figure 3 - CDF code distribution on EGEE using parrot software

## 4. CDF GRID USAGE, FUTURE PLANS AND CONCLUSIONS

During the past year CDF has been actively using both the EGEE and OSG grids for its scientific computing. Between July 2007 and June 2008 the CDF Virtual Organization (VO) CPU time used 1.4% of the CPU time used by all High Energy Physics VO's on EGEE grid. The CDF VO CPU usage on the Italian sites was 8.7% of CPU time used by all VO's. This is the largest among the non-LHC experiments. On the OSG the CDF CPU usage for the same time period was 46 Billion CPU hours. This corresponds to 18% of the CPU time used by the CMS (36%), US Atlas (27%), D Zero (20%) and CDF VO's (18%).

CDF has plans for future improvements to its grid computing. The glidein based job submission will migrate to the glideinWMS software [9]. GlideinWMS is a generic pilot-based workload management system based on the Condor glidein concept. GlideinWMS is being used by the CMS experiment. CDF specific extensions to the glideinWMS have been tested already. The use of glideinWMS makes it easier for CDF to exploit the opportunistic computing on the OSG. With glideinWMS, it is easier for the CDF add more grid sites to its NAMCAF portal and generate simulated data across more of the OSG than it currently does now. Along with opportunistic computing, CDF is working to make use of opportunistic storage in both the EGEE and OSG computing grids. The CDF computing portals will be modified.

CDF computing continues to be very successful as a result of the use of grid computing. Each computing grid (EGEE and OSG) required different solutions to fulfill CDF's computing requirements. Some common solutions are used (like Squid proxy Cache) and other solutions used on one grid will be migrated to the other grid (for example use of parrot to deliver the CDF software to the worker nodes). The use of computing portals is very important to the CDF experiment. Through their use, CDF has been able to transition to the grid with little or no impact on the Physicists analyzing the data. As future grid technologies are developed and standards arise CDF computing will evolve to meet the needs of the experiment.

### Acknowledgments

We thank the Fermilab staff and the technical staffs of the participating institutions for their vital contributions. This work was supported by the U.S. Department of Energy and National Science Foundation; the Italian Istituto Nazionale di Fisica Nucleare; the Ministry of Education, Culture, Sports, Science and Technology of Japan; the Natural Sciences and Engineering Research Council of Canada; the National Science Council of the Republic of China; the Swiss National Science Foundation; the A.P. Sloan Foundation; the Bundesministerium für Bildung und Forschung, Germany; the Korean Science and Engineering Foundation and the Korean Research Foundation; the Science and Technology Facilities Council and the Royal Society, UK; the Institut National de Physique Nucleaire et Physique des Particules/CNRS; the Russian Foundation for Basic Research; the Ministerio de Ciencia e Innovación, and Programa Consolider-Ingenio 2010, Spain; the Slovak R&D Agency; and the Academy of Finland.

### References

[1] dCache Project - (Deutsches Elektronen-Synchrotron, DESY and Fermi National Accelerator Laboratory, FNAL) – http://www.dcache.org
[2] M. Neubauer, "Computing for Run II at CDF", NIMA 502(2003) 386

[3] Condor Project – High Throughput Computing – http://www.cs.wisc.edu/condor/
[4] Open Science Grid – http://www.opensciencegrid.org
[5] Enabling Grids for E-SciencE (EGEE) - http://www.eu-egee.org/
[6] G. Compostella, D. Lucchesi, S.P. Griso, S.P., I.Sfligoi, "CDF Monte Carlo Production on LCG Grid via LcgCAF Portal" IEEE Inter. Conf. on e-Science and Grid Computing. Pages 11-16 Digital Object Identifier 10.1109/E-SCIENCE.2007.18
[7] Squid Web Proxy – http://www.squid-cache.org/
[8] D. Thain and M. Livny, "Parrot: An Application Environment for Data-Intensive Computing", Scalable Computing: Practice and Experience, Volume 6, Number 3, Pages 9--18, 2005.
[9] glideinWMS – Glidein Based WMS (Workload Management System) – http://www.uscms.org/SoftwareComputing/Grid/WMS/glideinWMS/